\documentclass[aps,twocolumn,prd,showpacs,showkeys,preprintnumbers,superscriptaddress,nobibnotes,floatfix,longbibliography,notitlepage,nofootinbib]{revtex4-1}

\pdfoutput=1
\usepackage{amsmath}
\usepackage{amsfonts}
\usepackage{amssymb}
\usepackage{mathrsfs}
\usepackage{graphicx}
\usepackage{xcolor}
\usepackage{xspace}
\usepackage{placeins}
\usepackage{fontawesome}

\usepackage{array}
\usepackage{bigints}
\usepackage{tabularx}

\usepackage[colorlinks]{hyperref}
\hypersetup{
    colorlinks,
    linkcolor={red!50!black},
    citecolor={blue!50!black},
    urlcolor={blue!80!black}
}
\usepackage{multirow}
\usepackage{upgreek}
\usepackage[capitalise]{cleveref}
\usepackage{soul}

\newcommand{\eq}{Eq.}
\newcommand{\eqs}{Eqs.}

\newcommand{\fig}{Fig.}

\newcommand{\figs}{Figures}
\newcommand{\Figs}{Figures}

\newcommand{\Treh}{T_{\mathrm{RH}}\xspace}

\newcommand{\Mbh}{M_{\bullet}}

\newcommand{\rbh}{r_h}
\newcommand{\Tbh}{T_H}

\newcommand{\ndm}{n_{dm}}
\newcommand{\Mpl}{M_{\mathrm{Pl}}}
\newcommand{\Mmin}{M_\mathrm{min}}

\begin{document}

\title{Dark matter from hot big bang black holes}

\author{Avi Friedlander}
\email{avi.friedlander@queensu.ca}
\affiliation{Department of Physics, Engineering Physics and Astronomy, Queen's University, Kingston ON K7L 3N6, Canada}
\affiliation{Arthur B. McDonald Canadian Astroparticle Physics Research Institute, Kingston ON K7L 3N6, Canada}

\author{Ningqiang Song}
\email{songnq@itp.ac.cn}
\affiliation{Institute of Theoretical Physics, Chinese Academy of Sciences, Beijing, 100190, China}
\affiliation{Department of Mathematical Sciences, University of Liverpool, \\ Liverpool, L69 7ZL, United Kingdom}

\author{Aaron C. Vincent}
\email{aaron.vincent@queensu.ca}
\affiliation{Department of Physics, Engineering Physics and Astronomy, Queen's University, Kingston ON K7L 3N6, Canada}
\affiliation{Arthur B. McDonald Canadian Astroparticle Physics Research Institute, Kingston ON K7L 3N6, Canada}
\affiliation{Perimeter Institute for Theoretical Physics, Waterloo ON N2L 2Y5, Canada}


\begin{abstract}
If the temperature of the hot thermal plasma in the Early Universe was within a few orders of magnitude of the Planck scale $\Mpl$, then the hoop conjecture predicts the formation of microscopic black holes from particle collisions in the plasma. Although these evaporated instantly, they would have left behind a relic abundance of all stable degrees of freedom which couple to gravity. Here we show that, upon minimal assumptions of a high reheat temperature and semiclassical black hole dynamics, this process could have produced the relic abundance of dark matter observed today for a particle mass anywhere in the range of $100~\mathrm{keV} \lesssim m_{dm} < M_{\rm Pl}$, though it could be subdominant to graviton-mediated freeze-in above $m_{dm} \sim$ MeV. The production mechanism does not rely on any additional assumptions about non-gravitational dark matter-Standard Model interaction.
\end{abstract}

\maketitle

\textbf{\textit{Introduction}} --- Big Bang cosmology describes an expanding universe originating from an initially hot, dense state. Its modern incarnation, the $\Lambda$CDM (dark energy-cold dark matter) \textit{concordance model}, accurately predicts nucleosynthesis and recombination, and with only six free parameters, is able to reproduce the observed power spectrum of the Cosmic Microwave Background (CMB) and Large Scale Structure~\cite{Planck:2018vyg}. However, $\Lambda$CDM remains agnostic about the origins of these free parameters, and about the initial temperature of the Universe, $\Treh$---usually called the \textit{reheating temperature}---so long as it is a factor of a few higher than what is necessary for nucleosynthesis.

Most attempts to explain the relic abundance of dark matter, $\Omega_{dm}h^2$, postulate some non-gravitational portal between the Standard Model and the dark sector,  that could lead to signatures in the laboratory from scattering, annihilation or production at colliders (see Refs.~\cite{Lisanti:2016jxe,Boveia:2018yeb} for reviews). However, as the evidence for dark matter comes via its gravitational effects alone, such a portal is not required. For instance, superheavy dark matter could have been produced from the vacuum fluctuations during or at the end of inflation~\cite{Chung:1998zb,Kuzmin:1998kk,Bhoonah:2020oov} or from Standard Model annihilation via gravitons~\cite{Tang:2017hvq,Garny:2017kha,Clery:2021bwz}. 

Here, we show that if the Universe started at a very high temperature, within a few orders of magnitude of the Planck mass $\Mpl$, the high-energy tail of the plasma's phase space distribution would lead to super-Planckian collisions which produce microscopic black holes. Because Hawking radiation produces all kinematically-accessible particles that couple to gravity, the subsequent rapid black hole evaporation  can yield the measured value of $\Omega_{dm}h^2$ with minimal assumptions, and without requiring the dark sector to communicate non-gravitationally with the Standard Model.

The production of dark matter from black hole evaporation has been examined in Refs. \cite{Matsas:1998zm,Bell:1998jk,Khlopov:2004tn,Fujita:2014hha,Allahverdi:2017sks,Lennon:2017tqq,Morrison:2018xla,Hooper:2019gtx,Dalianis:2019asr,Chaudhuri:2020wjo,Hooper:2020evu,Masina:2020xhk,Baldes:2020nuv,Keith:2020jww,Gondolo:2020uqv,Cai:2020kfq,Hooper:2020otu,Bernal:2020kse,Bernal:2020ili,Bernal:2020bjf,Auffinger:2020afu,Kitabayashi:2021hox,Masina:2021zpu,Arbey:2021ysg,Baker:2021btk,Schiavone:2021imu,Sandick:2021gew,Calza:2021czr,Smyth:2021lkn,Barman:2021ost,Samanta:2021mdm,Kitabayashi:2022fqq,Barman:2022gjo,Cheek:2022dbx,Kitabayashi:2022uvu,Mazde:2022sdx,Barman:2022pdo,Bhaumik:2022zdd,Cheek:2022mmy,Agashe:2022phd,Marfatia:2022jiz,Chaudhuri:2023aiv,Cheek:2021cfe,Cheek:2021odj}. Those scenarios rely on a well-established but ``beyond-$\Lambda$CDM'' black hole production mechanism, such as non-Gaussian field excursions during inflation. Alternatively, the production and evolution of primordial black holes from particle collisions has been studied in scenarios with exotic assumptions about the Planck scale~\cite{Guedens:2002sd,Majumdar:2002mra,Sendouda:2003dc,Sendouda:2004hz,Tikhomirov:2005bt,Conley:2006jg,Dubrovich:2021gdn,Friedlander:2022ttk}. 

In contrast to previous approaches, the scenario that we examine here only makes minimal assumptions about standard Big Bang cosmology, namely: 1) a high $\Treh$ and 2) semiclassical arguments about the formation and evaporation of black holes remain valid near the Planck scale. These $\Treh$ values are larger than allowed by the simplest slow roll single field inflation models. However, no model independent constraints exist on such $\Treh$ values. The first upper bounds on $\Treh \sim \Mpl$ should come from next-generation CMB observatories~\cite{Ringwald:2020ist}. 

While little is conclusively known about the breakdown of semiclassical black hole physics, we will present results for different assumptions on the impact of quantum gravity on Planck-scale black holes including a demonstration that the qualitative results in this work hold even if the formed black holes are restricted to a regime where semiclassical arguments are expected to be trustworthy. 

We will show that dark matter can be produced gravitationally in the early Universe without requiring any non-gravitational interactions or exotic cosmology beyond the $\Lambda$CDM framework. While the absence of non-gravitational interactions would make direct detection challenging, there are observational signatures that arise from the requirement of a hot big bang. Furthermore, black hole evaporation yields a non-thermal dark matter velocity distribution. We will briefly return to these ideas in the final sections.

Hereafter, we work in ``particle physicist units'' where $c=\hbar=k_B =1$, but $\Mpl$ carries units of energy.

\textbf{\textit{Formation and evaporation of black holes in the primordial plasma}} --- A black hole of mass $\Mbh$ is characterized by its horizon radius $\rbh(\Mbh)$. The hoop conjecture posits that when two particles with centre-of-mass energy $\sqrt{s}$ come within an impact parameter $2r_h(\sqrt{s})$ of each other, they will form a black hole of mass $\Mbh = \sqrt{s}$~\cite{thorne1972nonspherical,Song:2019lxb}. If the collision is head-on and has no net charge, then $\rbh = {2 \Mbh}/{\Mpl^2}$ and a neutral, non-rotating Schwartszchild black hole is formed. In reality, particle collisions will not be exactly head-on and many would have non-zero net charge. However, the majority of formed black holes would be non-extremal. While charged and rotating black holes would have altered evaporation rates leading to $\mathcal{O}(1)$ changes in the relic abundance, these  are unlikely to affect our overall conclusions.

Even if the plasma temperature is well below the Planck scale, the high-energy tail of the phase space distribution can extend to energies above $\Mpl$. During radiation domination, the black hole collisional production rate per unit mass per unit volume is~\cite{Friedlander:2022ttk,SupplementalMaterial}
\begin{equation}
    \dfrac{d\Gamma}{d\Mbh}=\dfrac{g_{\star}(T)^2 \Mbh^5 T^2}{2\pi^3 \Mpl^4}\left[\dfrac{\Mbh}{T}K_1(\frac{\Mbh}{T})+2K_2(\frac{\Mbh}{T})\right],
    \label{eq:BHprodrate}
\end{equation}
as long as $\Mbh$ is greater than $\Mmin$, the minimum mass-energy required to form a black hole. Here, $g_*(T)$ is the  effective number of relativistic degrees of freedom in the plasma, which has a temperature $T$, and $K_i(x)$ is the modified Bessel function of the second kind. A derivation of Eq.\eqref{eq:BHprodrate} can be found in the Spp. Mat.~\ref{app:BHProdRate}. Because $K_i(x) \approx \sqrt{\pi/(2x)}e^{-x}$ for large $x$, in the limit of temperature $T\ll \Mbh$ the black hole production rate is approximately
\begin{equation}
    \dfrac{d\Gamma}{d\Mbh}=\dfrac{g_{\star}(T)^2 \Mbh^{11/2} T^{3/2}}{2\sqrt{2}\pi^{5/2} \Mpl^4}e^{-\Mbh/T}\,,
    \label{eq:BHprodrateLowT}
\end{equation}
for $\Mbh> \Mmin$. A simple assumption is that $\Mmin\simeq \Mpl$. However, in the absence of a theory of quantum gravity, arguments can be made about the minimum black hole mass being larger or smaller than $\Mpl$. We will thus explore a range of $\Mmin$ values, and will find that it does not qualitatively affect our conclusions as long as it is within an order of magnitude of the Planck mass.

The black holes thus formed will evaporate nearly instantaneously. Hawking's semiclassical arguments predict a distorted thermal spectrum comprising all degrees of freedom that couple to gravity. The evaporation rate is characterized by a Hawking Temperature $T_H = \Mpl^2/8\pi\Mbh$~\cite{Hawking:1975vcx}. 

The rate of mass loss from evaporation to species $i$ can be compactly expressed as
\begin{equation}
    -\frac{d M_{\bullet\to i}}{dt} = \frac{g_{i}}{2\pi^2}\int \dfrac{E \sigma_i(r_hE)}{\exp(E/T_H)\mp 1}p^2dp\,,
    \label{eq:dMdtSpecies}
\end{equation}
where the $(+)$ sign applies to fermions and the $(-)$ sign to bosons. $g_i$ is the number of degrees of freedom in species $i$, and $E$ and $p$ are respectively the energy and momentum of the evaporation product. The greybody distortion factors $\sigma_i$ encode the probability of a particle escaping to ``infinity'', and are computed via partial-wave scattering. Here, we use values provided by \texttt{BlackHawk} \cite{Arbey:2019mbc,Arbey:2021mbl} and compiled in \texttt{CosmoLED}~\cite{Friedlander:2022ttk}.

Going forward, we take the limit where the masses of the evaporation products  $m_i$ are much lower than $T_{H}$, which will hold for $m_i \ll \Mpl$.  \eq~\eqref{eq:dMdtSpecies} then integrates to a constant times $T_{H}^2$, and the total evaporation rate is a sum over all evaporation species:
\begin{equation}
    \frac{d M_{\bullet}}{dt} = \sum_i \dfrac{d M_{\bullet\to i}}{dt}=-\sum_ig_i \alpha_i \Tbh^2 = -\alpha_{\rm tot} T_H^2,\,
    \label{eq:dMdtEvap}
\end{equation}
where the $\alpha_i$ are constants that depend only on the spin of species $i$. These are provided in Tab.~\ref{tab:alphaBetaVals} for spin 0, 1/2 and 1 particles, along with $\alpha_{tot}$, which is simply a sum of the $\alpha_i$ weighted by the number of degrees of freedom in each sector in the Standard Model. Graviton (spin-2) production is suppressed, and does not affect the value of $\alpha_{tot}$ to the precision that we quote.

\begin{table}[!t]
\caption{Values of $\alpha_i$ and $\beta_i$ for each particle spin. These values describe how black holes evaporate in the high-temperature limit as defined in \eqs~\eqref{eq:dMdtEvap} and \eqref{eq:dNdt}.}
\begin{center} 
\begin{tabular}{c  c c  c  c } 
 \hline \hline
  & Scalar & Spinor & Vector &  Standard Model \\ [0.5ex] 
 \hline
 $\alpha_i$ & $4.67\times 10^{-2}$  & $2.58\times 10^{-2}$ & $1.06\times 10^{-2}$  & $\alpha_{tot}= 2.77$ \\ 
 $\beta_i$ & $1.67\times 10^{-2}$ & $6.10\times 10^{-3}$ & $1.86\times 10^{-3}$ &  -- \\
 \hline \hline
\end{tabular}
\end{center}

\label{tab:alphaBetaVals}
\end{table}
The \textit{number} of particles produced is found in a near-identical way to Eqs. (\ref{eq:dMdtSpecies}-\ref{eq:dMdtEvap}), but dropping the factor of $E$ in the integrand of Eq. \eqref{eq:dMdtSpecies}. At high $T_{H}$, this yields a production rate of particles of species $i$ per black hole:
\begin{equation} \label{eq:dNdt}
\frac{dN_i}{dt} = g_i \beta_i \Tbh,
\end{equation}
where the $\beta_i$ are factors that again depend only on the spin; these are given in Tab. \ref{tab:alphaBetaVals}. 
The total number of particles of each species produced by a black hole of initial mass $\Mbh$ is thus
\begin{equation}
    N_{i} = \int_{\Mbh}^0 d\tilde{\Mbh}   \frac{dN_i}{dt}  \bigg(\frac{d \tilde{\Mbh}}{dt}\bigg)^{-1} = \frac{4\pi g_{i}  \beta_i \Mbh^2}{\alpha_{\rm tot}\Mpl^2 }\,.
\end{equation}
The majority of particle production occurs while the black hole is within an $\mathcal{O}(1)$ factor of its starting mass. Therefore, we can take the integral's lower mass limit to zero without concern for how quantum gravity impacts the evaporation of black holes with $\Mbh \ll \Mpl$. Our results with a larger $\Mmin$ value act as a conservative estimate if only evaporation above the Planck scale is trusted.

\textbf{\textit{Relic abundance of dark matter}} --- We now specialize to the case of interest, the production of a stable dark matter species with $g_{dm}$ degrees of freedom and mass $m_{dm} \ll \Mpl$ that is thermally decoupled from the primordial plasma. 

Dark matter production shuts off exponentially as the Universe cools, so the number density of dark matter particles at $T = T_{RH}$ is found, to good accuracy, by convolving the black hole production rate with $N_{dm}$ over all temperatures:

\begin{equation} \label{eq:ndmImplicit}
    n_{dm}(\Treh) \approx \int_{\Treh}^0 \!\!\!dT\bigg(\frac{dT}{dt}\bigg)^{-1}  \!\!\int_{\Mmin}^\infty d\Mbh N_{dm} \dfrac{d\Gamma}{d\Mbh} \,.
\end{equation}
During radiation domination, $dT/dt=-HT$ and the Hubble parameter is given by the Friedmann equation, $H^2=8\pi\rho_r/(3\Mpl^2)$, where the radiation density is $\rho_r=\pi^2g_* T^4 / 30$.
The integral in \eq~\eqref{eq:ndmImplicit} can be performed analytically, yielding
\begin{equation}\label{eq:ndmTRH}
n_{dm}(\Treh)\approx \frac{3\sqrt{\frac{5}{2}} g_*(\Treh)^{3/2} g_{dm} \beta_{dm}}{\pi^3\alpha_{\rm tot}} \frac{\Mmin^8}{\Mpl^5} f_{dm}\bigg(\frac{T_{RH}}{\Mmin}\bigg),
\end{equation}
where the dependence on $\Treh$ is encoded by the function $f_{dm}(x)$. To lowest order in $\Treh/\Mmin$, this is
\begin{equation}\label{eq:ndmTRHFirstOorder}
f_{dm}\bigg(\frac{T_{RH}}{\Mmin}\bigg) \simeq    \left(\frac{\Treh}{\Mmin}\right)^{3/2} e^{-\Mmin/\Treh} \,. 
\end{equation}
Going forward, we will nonetheless use the full (but less intuitive) analytic result of the integral in Eq.~\eqref{eq:ndmImplicit}. The full expression as well as a comparison with Eq. \eqref{eq:ndmTRHFirstOorder} can be found in Supp. Mat. \ref{app:validity}.

Finally, if the dark sector is fully decoupled, it only redshifts due to adiabatic expansion, and the conservation of entropy yields a present-day dark matter density:
\begin{equation} \label{eq:OmDM}
    \Omega_{dm} = \frac{g_{*s}(T_0) T_0^3 n_{dm}(\Treh) m_{dm}}{g_{*s}(\Treh) \Treh^3 \rho_{c}}\,,
\end{equation}
where $g_{*s}$ is the effective number of degrees of freedom contributing to entropy, $T_0$ is the CMB radiation temperature today, and $\rho_{c}$ is the present-day critical density. The abundance of DM thus scales as:
\begin{equation}
    \Omega_{dm} h^2 \sim 0.1 \, \frac{g_{dm}m_{dm}}{10\,{\rm keV}} \left(\frac{\Mmin}{\Mpl}\right)^5 \left(\frac{\Mmin}{10\,\Treh}\right)^{3/2}\! \! \! e^{-\frac{\Mmin}{10\,\Treh}},
\end{equation}
as long as $g_{dm}$ is significantly smaller than the total number of Standard Model degrees of freedom.

\begin{figure}[ht]
    \centering\includegraphics[width=0.5\textwidth]{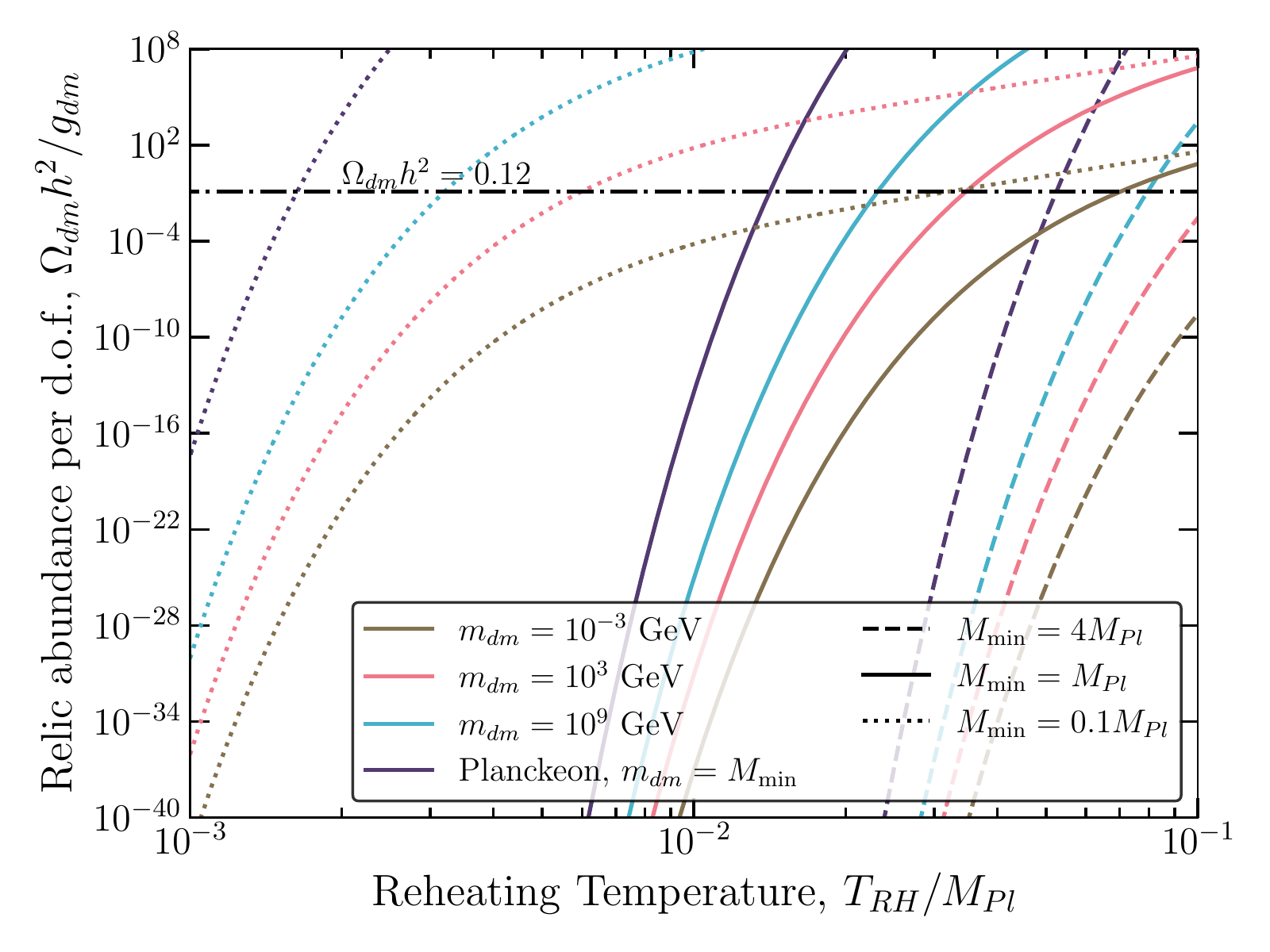}
    \caption{\textbf{\textit{
    Dark matter relic abundance per degree of freedom  produced from microscopic black hole evaporation.}} Line colours depict fermionic dark matter with different masses $m_{dm}$ or a Planckeon with mass $\Mmin$. For each dark matter we show three production scenarios: The standard assumption on the minimum black hole mass $\Mmin=\Mpl$, an optimistic $\Mmin=0.1\Mpl$ where microscopic black holes are more efficiently produced, and the conservative assumption $\Mmin = 4 \Mpl$. The dot-dashed horizontal line shows the observed abundance of dark matter $\Omega_{dm}h^2=0.12$ with $g_{dm}=1$~\cite{Planck:2018vyg}. For all scenarios, microscopic black holes are produced by collisions in the Standard Model plasma with $g_*(\Treh)=g_{*s}(\Treh)=106.75$.
    }
    \label{fig:OmDMZoomOut}
\end{figure}

\fig~\ref{fig:OmDMZoomOut} shows the relic abundance of dark matter with different masses as a function of the reheating temperature. Because the Hawking temperature is much higher than the dark matter masses we are considering, production is relativistic and the process leads to a fixed number density for a given $\Mmin$ and $\Treh$. It is thus insensitive to the dark matter mass except for the linear scaling in \eq~\eqref{eq:OmDM}. A larger $\Treh/\Mmin$ predictably yields more black hole production, and more transfer of energy into the dark sector. 

\begin{figure}[ht]
    \centering
    \includegraphics[width=0.5\textwidth]{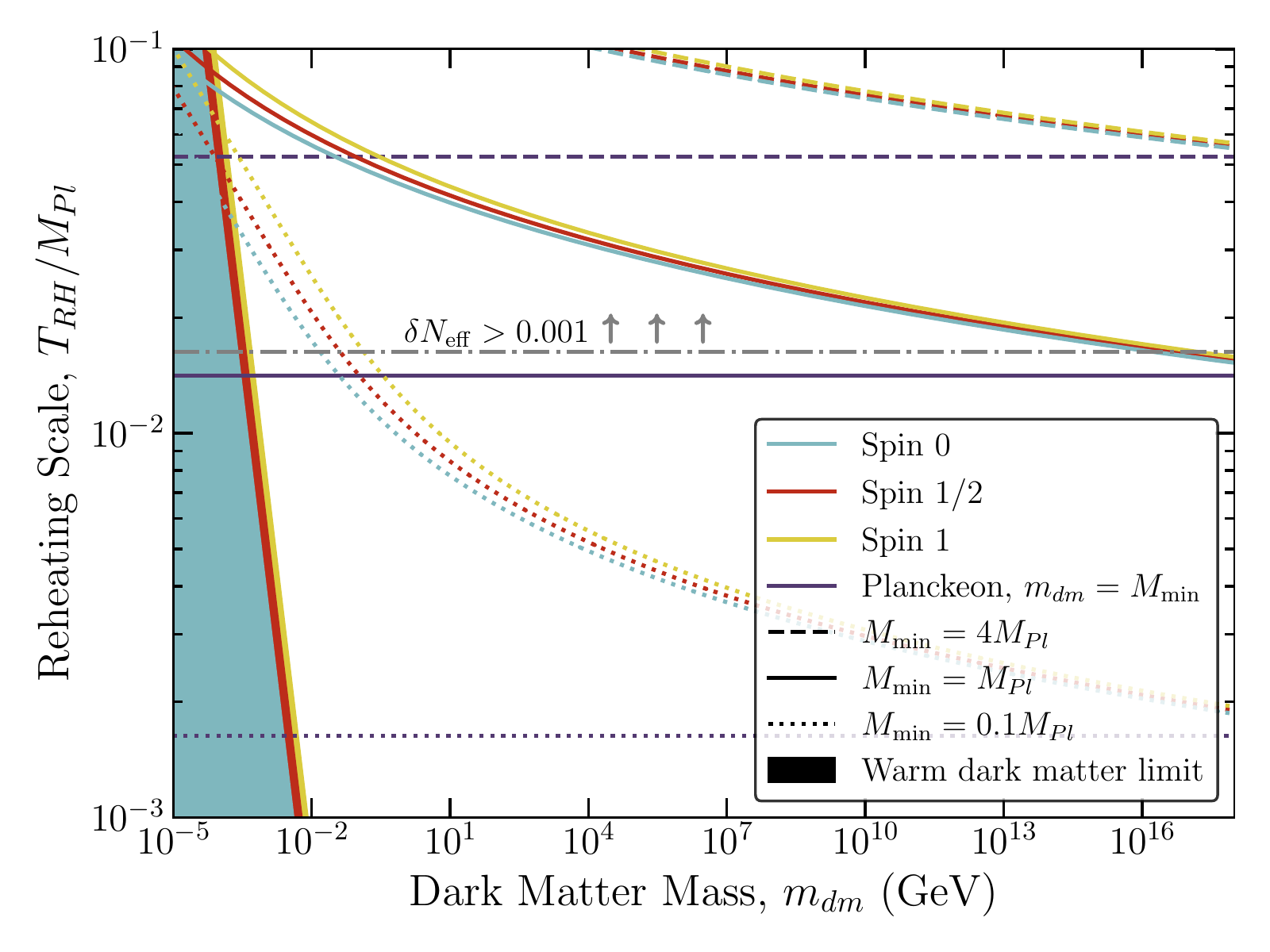}
    \caption{\textbf{\textit{Required reheating temperature to produce the observed relic abundance of dark matter today.}} Blue, red, and yellow curves show reheating temperature required to produce scalar, spinor, or vector dark matter respectively with $g_{dm}=1$. Horizontal purple lines show the required reheating temperature to produce Planckeon dark matter if $\Mmin$ mass black holes are stable. Different line styles show different assumptions of the minimum black hole mass, $\Mmin$.The shaded region is excluded from warm dark matter constraints, and the gray horizontal dot-dashed line depicts shows $\Treh$ above which $N_{\rm eff}$ is modified beyond its theoretical uncertainty. }
    \label{fig:TRHdm}
\end{figure}

In the absence of a theory of quantum gravity, there is significant uncertainty about Planck scale black hole production and evaporation. \Figs~\ref{fig:OmDMZoomOut} and \ref{fig:TRHdm} therefore present results for a range of values of $\Mmin$ from 0.1 to 4 times the Planck mass, covering optimistic to conservative scenarios. At the higher end of this range, we have demanded that the evaporation timescale be much longer than the horizon formation rate, which amounts to a condition on the black hole entropy, and ensures that semiclassical assumptions remain valid \cite{Mack:2019bps}. However, sub-Planckian black holes have been discussed in the context of certain quantum gravity frameworks \cite{Carr:2015nqa,Modesto:2009ve}, and it is furthermore not clear that horizon-formation is even necessary for this mechanism to have produced our dark matter, as pre-Hawking radiation which prevents total collapse could have very similar consequences as seen by an observer at ``infinity'' \cite{Vachaspati:2006ki,Vachaspati:2007hr,Dai:2007ki}. 

\fig~\ref{fig:TRHdm} shows the locations in the $\Treh$--$m_{dm}$ plane that yield the observed abundance of dark matter today. Different line styles correspond to different assumptions for the minimum allowed black hole mass, while colours correspond to dark matter spin. Because of the exponential scaling with $\Treh$, dark matter with masses from a few keV to $\gtrsim 10^{18}$ GeV can be produced in the observed amount within only a comparatively small window of reheating temperatures $10^{-3} \lesssim \Treh/\Mpl \lesssim 0.1$. A higher minimum mass for black hole formation predictably raises the required $\Treh$.

For completeness we also consider the possibility that the evaporating black hole ends up as a stable remnant with $\Mbh\sim \Mmin$, the so called \textit{Planckeon}\footnote{Also called  ``Planck Relics'' or ``Black Hole Remnants'', see Ref.~\cite{Chen:2014jwq} for a review.} \cite{Aharonov:1987tp}, which would act as cold dark matter, an alternative channel to evaporative production. Planckeon production by particle collisions during a hot big bang has been previously proposed~\cite{Barrau:2019cuo}; we improve on their estimates by making use of  the full differential black hole production rate, discussed above~\cite{SupplementalMaterial}. We show in \figs~\ref{fig:OmDMZoomOut} and~\ref{fig:TRHdm} the Planckeon relic abundance and the corresponding reheating temperatures. In \fig~\ref{fig:TRHdm}, the Planckeon lines are horizontal because their mass is held fixed at $m_{dm}=\Mmin$. These figures demonstrate that if Planckeons are stable, their abundance will always dominate over dark matter species produced from evaporation. Since there is no exact prediction for the Planckeon mass (it should be near the Planck mass, but not necessarily equal to it due to quantum gravity effects), we show a range of $M_{\mathrm{min}}$ values.   In addition, black holes could be sub-Planckian in certain models \cite{Carr:2015nqa,Modesto:2009ve}. We note that if a significant fraction of Planckeons retains some electric charge their presence as the dominant dark matter component may be discovered or severely constrained in the near future~\cite{Lehmann:2019zgt}.

\textbf{\textit{Warm Dark Matter Constraints}} --- The matter power spectrum is measured via the Lyman-$\alpha$ forest and places strong constraints on warm dark matter (WDM). This can be translated to a lower limit on the mass of dark matter from black hole evaporation. Because evaporation produces a distorted thermal spectrum, we follow the method of Ref. \cite{Baldes:2020nuv}, and demand that the average dark matter velocity is at most the average velocity of WDM \cite{Bode:2000gq} with the minimum mass allowed by Lyman-$\alpha$ observations, $m_{\rm min}^{wdm}$.

We obtained the average dark matter momentum at production by reproducing the procedure from previous sections, finding the total energy density of dark matter produced, $\rho_{dm}$, and noting that for relativistic particles the average momentum is simply $\rho_{dm}/n_{dm}$. If the dark matter momentum redshifts only due to Universe expansion, this yields a limit on $m_{dm}$:
\begin{equation}
    m_{dm} \gtrsim 10^{-5}\,\mathrm{ GeV} \frac{\alpha_{dm}}{\beta_{dm}} \left( \frac{0.1 \Mpl}{\Treh} \right) \left( \frac{m_{\rm min}^{wdm}}{4\,\mathrm{ keV}}\right)^{4/3} \,,
\end{equation}
where we take the WDM mass limit of $m_{\rm min}^{wdm} = 4.65$~keV~\cite{Yeche:2017upn}. These  limits are shown as a shaded region in \fig ~\ref{fig:TRHdm}, and constrain dark matter produced in black hole evaporation to be heavier than $\sim 100$ keV.

\textbf{\textit{Large Reheating Temperature}} ---
In single-field slow-roll inflationary models, CMB meaasurements of the scalar power spectrum amplitude $A_s$ and limits on the tensor-to-scalar ratio $r$ set an upper limit on the energy density at the end of inflation of $\rho_{RH} \lesssim (1.6\times10^{16}\mathrm{~ GeV})^4$~\cite{Planck:2018jri}. However, these do not directly apply to some multi-field inflation models \cite{Byrnes:2006fr} or to non-inflationary theories of the early Universe such as a cyclic Universe \cite{Boyle:2003km} or String Gas Cosmology \cite{Brandenberger:2006xi}.

Because the mechanism discussed here can in principle produce a relic abundance of completely sterile dark matter, the high $\Treh$ required is one of the few model-independent observational targets. A large reheating temperature would enhance the normally $\Mpl^{-2}$-suppressed background of gravitational waves with frequencies $\mathcal{O}(100~\mathrm{ GHz})$ \cite{Ghiglieri:2015nfa, Ghiglieri:2020mhm, Ringwald:2020ist, Ghiglieri:2022rfp}. These   would contribute to the effective number of relativistic degrees of freedom at recombination, $N_{\rm eff}$. While current $N_{\rm eff}$ measurements do not constrain sub-Planckian $\Treh$ values, the next generation of CMB experiments could probe $\Treh \lesssim \Mpl$ \cite{Ringwald:2020ist}. Ultimately, if CMB experiments are able to constrain $N_{\rm eff}$ down to the theoretical uncertainty limit $\delta N_{\rm eff} \sim 10^{-3}$, then $\Treh < 2\times10^{17}$~GeV could be within reach~\cite{Ghiglieri:2020mhm}, conclusively testing the microscopic black hole production of dark matter with $\Mmin \geq \Mpl$ and $m_{dm} \lesssim 10^{16}$~GeV as seen by the gray dot-dashed line in \fig~\ref{fig:TRHdm}. The high-frequency gravitational waves from $\Treh$ and from black hole evaporation could be converted to microwaves in a strong magnetic field via the inverse Gertsenshtein effect~\cite{gertsenshtein1962wave}. However, even optimistic predictions lead to similar sensitivities~\cite{Ringwald:2020ist} as $N_{\rm eff}$, and the technology remains largely undeveloped.

\textbf{\textit{ Freeze-in Via Gravitons}} --- During a hot big bang, purely gravitational dark matter may also be produced by Standard Model paticles annihilating via a graviton portal \cite{Tang:2017hvq,Garny:2017kha,Clery:2021bwz}. If this portal exists and, if the tree-level graviton exchange diagram is reliable, then dark matter with a mass of $m_{dm} \sim 10^{12}$ GeV can be produced with a reheating scale as low as $\Treh\lesssim10^{-6} \Mpl$~\cite{Tang:2017hvq}. In this case, black hole production remains dominant for masses $m_{dm}\lesssim 1$ MeV assuming instantaneous reheating. These two mechanisms rely on different assumptions about quantum gravity: a reliable tree-level non-renormalizable low-energy effective quantum theory for freeze-in, described by the Lagrangian $\mathcal{L}=-\frac{1}{\Mpl}h_{\mu\nu}T^{\mu\nu}$ with $h_{\mu\nu}$ promoted to be the graviton propagator, versus the rubustness of the semiclassical regime at high energies for black hole production. The dependence on the inflation model may also differ (see e.g.~\cite{Clery:2021bwz}). Taken together, these two scenarios point at a dominant gravitational portal for dark matter production in the case of a high reheating scale that is robust against assumptions about the fundamental nature of quantum gravity.

\textbf{\textit{ Conclusions}} ---
We have presented a novel mechanism which allows for the gravitational production of dark matter that does not rely on any non-gravitational interactions with the Standard Model or specific inflationary physics. The only non-standard assumption required to achieve the observed relic abundance is a large $\Treh \gtrsim 10^{17}$~GeV. While the curves presented in \figs~\ref{fig:OmDMZoomOut} and \ref{fig:TRHdm} rely on only the existence of Standard Model particles and one dark matter species, additional stable or unstable particle species between the Weak and Planck scales do not affect our conclusions, and are only a matter of bookkeeping in the values of $g_{dm}$, $\alpha_{\rm tot}$, $g_\star$, and $g_{\star S}$ at $\Treh$.

A more detailed treatment including the effects of black hole rotation and charge will yield a more accurate prediction of the values of $\Treh$ and $\Mmin$ required to produce the dark matter observed today. As such greybody factors differ only by $\sim O(1-10)$, we do not expect a qualitative change in the overall conclusions. 

Observational handles on such a secluded sector produced at such early times remain few and far between. However, discovery of a large near-Planck scale $\Treh$ will set a model-independent upper bound on the mass of any new stable particle which does not reach chemical equilibrium with the Standard Model. Subtle gravitational-wave observables may hint at such high reheating temperatures, and quantifying the impact of this non-thermal dark matter distribution on large scale structure may provide additional clues. Ultimately, the question of dark matter's true nature may be tied to the very early Universe and quantum gravity, and if we can pull on a single thread, it may reveal far more than expected.

\section*{Acknowledgements}
We thank Joe Bramante and Sarah Shandera for helpful discussions.
AF is supported by an Ontario Graduate Scholarship. NS is supported by the National Natural Science Foundation of China (NSFC) Project No. 12047503. NS also  acknowledges the UK Science and Technology Facilities Council for support through the Quantum Sensors for the Hidden Sector collaboration under the grant ST/T006145/1. ACV is supported by the Arthur B.~McDonald Canadian Astroparticle Physics Research Institute, NSERC, and the province of Ontario via an Early Researcher Award. Equipment is funded by the Canada Foundation for Innovation and the Province of Ontario, and housed at the Queen's Centre for Advanced Computing. Research at Perimeter Institute is supported by the Government of Canada through the Department of Innovation, Science, and Economic Development, and by the Province of Ontario.
\bibliography{LEDBH.bib}
\clearpage
\newpage

\maketitle
\onecolumngrid
\begin{center}
\textbf{\large Dark matter from hot big bang black holes}

\vspace{0.05in}
{ \it \large Supplementary Material}\\ 
\vspace{0.05in}
{Avi Friedlander, Ningqiang Song and Aaron C. Vincent}
\end{center}
\onecolumngrid
\setcounter{equation}{0}
\setcounter{figure}{0}
\setcounter{section}{0}
\setcounter{table}{0}
\setcounter{page}{1}
\makeatletter
\renewcommand{\theequation}{S\arabic{equation}}
\renewcommand{\thefigure}{S\arabic{figure}}
\renewcommand{\thetable}{S\arabic{table}}
\vspace{-3mm}

\section{Black hole production rate} \label{app:BHProdRate}

In a high-temperature relativistic plasma, the collisional black hole formation rate per unit volume per unit mass can be written as~\cite{Conley:2006jg}
\begin{equation}
    \dfrac{d\Gamma}{d\Mbh}=g_{\star}(T)^2\int \dfrac{d^3k_1}{(2\pi)^3}\dfrac{d^3k_2}{(2\pi)^3}e^{-k_1/T}e^{-k_2/T}\sigma(\Mbh)v_{rel}\delta\left(\sqrt{(k^\mu_1+k^\mu_2)^2}-\Mbh\right)\Theta(\Mbh-\Mmin)\,,
    \label{supp_eq:dGammadMStart}
\end{equation}
where we have assumed that the thermal distribution of parent particles is approximately Maxwellian because particles with $k \ll T$ are irrelevant for black hole formation. The step function $\Theta$ ensures $E_{CM}\geq \Mmin$. The integral can be evaluated if $v_{rel}=|\vec{v}_1-\vec{v}_2|\simeq 1$ by defining $\theta$ as the angle between the incoming particles and integrating over all other angles. This results in:
\begin{equation}
    \dfrac{d\Gamma}{d\Mbh}=
    \frac{g_{\star}(T)^2 \sigma(\Mbh)}{8\pi^4}
    \int dk_1 dk_2 d\cos(\theta) e^{-k_1/T}e^{-k_2/T}
   k_1^2k_2^2 \delta\left(\sqrt{2k_1 k_2 (1-\cos\theta)}-\Mbh\right)\Theta(\Mbh-\Mmin)\,.
    \label{supp_eq:dGammadMIntermediate}
\end{equation}
These integrals can be performed analytically resulting in a formation rate of
\begin{equation}
    \dfrac{d\Gamma}{d\Mbh}=\dfrac{g_{\star}(T)^2\sigma(\Mbh) }{32\pi^4}\Mbh^3 T^2\left[\dfrac{\Mbh}{T}K_1(\frac{\Mbh}{T})+2K_2(\frac{\Mbh}{T})\right]\Theta(\Mbh-\Mmin)\,,
    \label{supp_eq:dGammadMSigmaDependent}
\end{equation}
where $K_i(x) $ is the modified Bessel function of the second kind. Finally, the black hole production rate is found to be
\begin{equation}
    \dfrac{d\Gamma}{d\Mbh}=\dfrac{g_{\star}(T)^2 \Mbh^5 T^2}{2\pi^3 \Mpl^4}\left[\dfrac{\Mbh}{T}K_1(\frac{\Mbh}{T})+2K_2(\frac{\Mbh}{T})\right]\Theta(\Mbh-\Mmin),
    \label{supp_eq:BHprodrate}
\end{equation}
by assuming that the production cross-section is equal to the geometric limit:
\begin{equation}
    \sigma(\Mbh)  = 4 \pi \rbh^2 \,.
\end{equation}

\section{Analytic expressions and their accuracy} \label{app:validity}

In deriving the relic abundance of dark matter we used three approximations:
\begin{enumerate}
    \item Using the $ T \ll \Mbh $ limit for the production rate of black holes. 
    \item Assuming black hole evaporation is independent of dark matter mass. 
    \item Presenting the analytic expressions only to lowest order in $\Treh/\Mmin$.
\end{enumerate}
The results shown in the main text use the first two approximations, whereas the third  is only used to present the analytic equation in a more intuitive form. In this section we demonstrate the range of validity of these approximations and present the full analytic form for dark matter number density.

\begin{figure}[htb]
    \centering
    \includegraphics[width=0.48\textwidth]{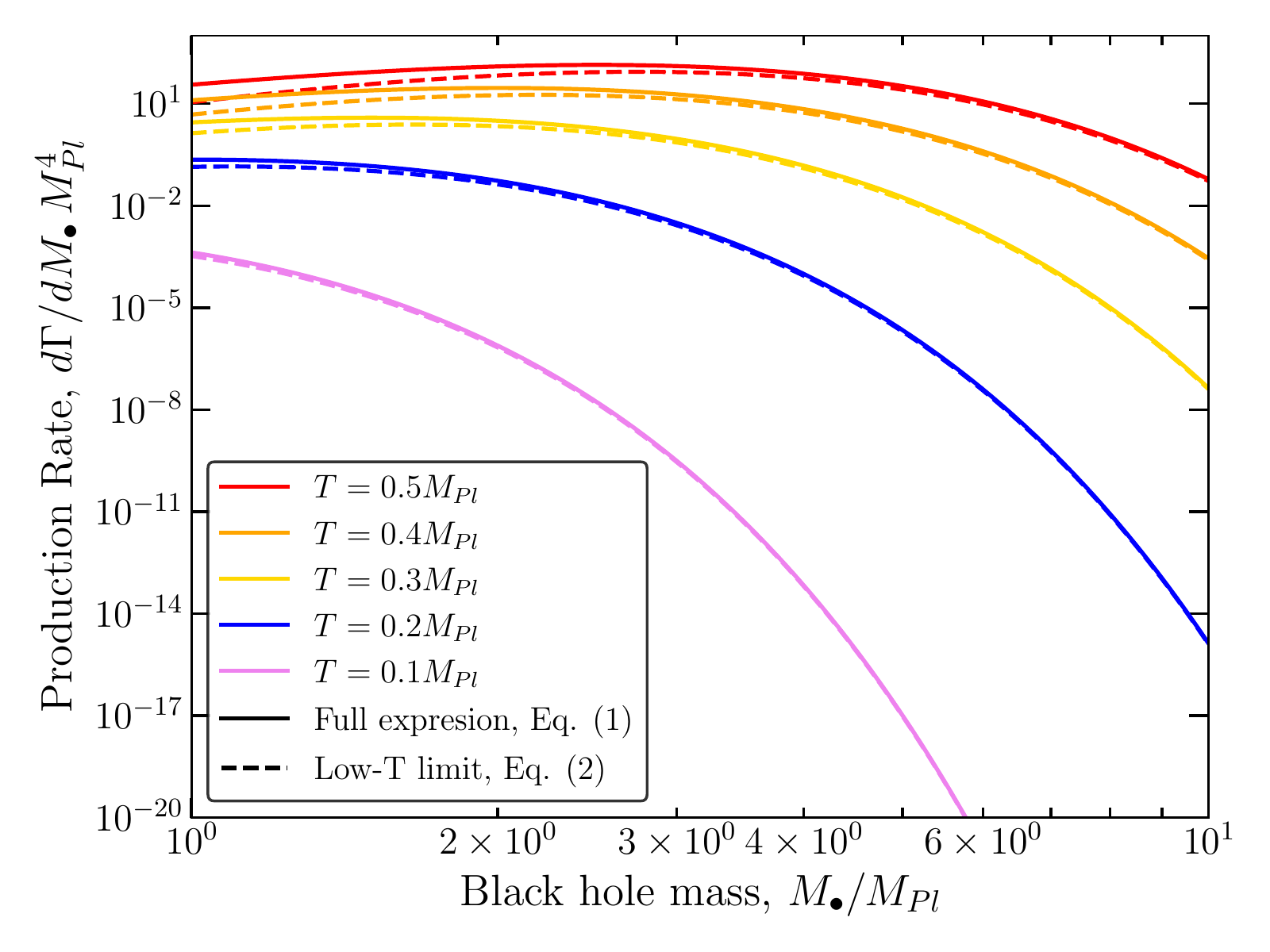}
    \includegraphics[width=0.48\textwidth]{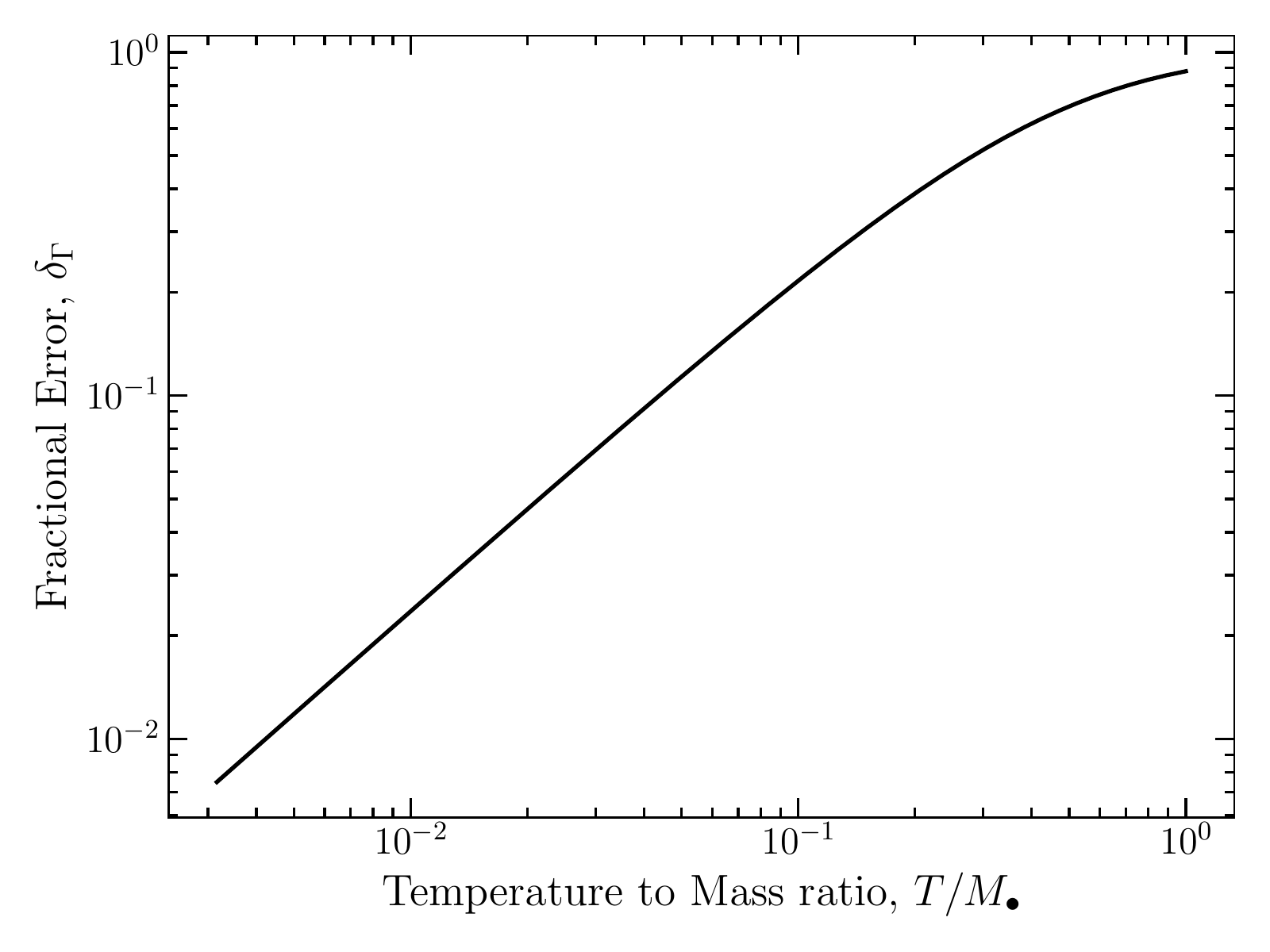}
    \caption{Left: The production rate of microscopic black holes in the very early Universe from collisions in the plasma. The solid curves depict the analytic expression \eqref{supp_eq:BHprodrate} found by integrating over the Maxwell-Boltzmann distributions of two incoming particles. The dashed curves show the production rate in the $T \ll \Mbh$ limit which is used to produce the results in this work. The different colour curves show the production rates at different plasma temperatures. Right: The fractional error that arises from using the low-$T$ limit for black hole production rate as a function of $T/\Mbh$. For both panels, the plasma is assumed to begin as only containing Standard Model particles so that $g_*(\Treh)=106.75$.}
    \label{fig:BHProdRate}
\end{figure}

For the first approximation, we simplified the production rate of black holes, $d\Gamma/d\Mbh$ by assuming that the plasma temperature, $T$, is much smaller than the mass of the formed black hole, $\Mbh$. In this limit, the Bessel function is $K_i(x) \approx \sqrt{\pi/(2x)}e^{-x}$ for large $x$ resulting in an approximate production rate of
\begin{equation}
    \dfrac{d\Gamma}{d\Mbh}\approx \dfrac{g_{\star}(T)^2 \Mbh^{11/2} T^{3/2}}{2\sqrt{2}\pi^{5/2} \Mpl^4}e^{-\Mbh/T}\,.
    \label{supp_eq:BHprodrateLowT}
\end{equation}
The left panel of \fig~\ref{fig:BHProdRate} shows the black hole production rate at different temperatures. The full expression which accounts for the integral over the full Maxwell-Boltzmann distribution is shown as solid curves, whereas the dashed curves show the low-T approximation. 

By directly comparing the approximation \eqref{supp_eq:BHprodrateLowT} to the full equation \eqref{supp_eq:BHprodrate}, the fractional error, $\delta_\Gamma$, from using this approximation is
\begin{equation}
    \delta_\Gamma = 1 - \sqrt{\frac{2 \Mbh}{\pi T}} e^{-\Mbh/T} \bigg(\dfrac{\Mbh}{T}K_1(\frac{\Mbh}{T})+2K_2(\frac{\Mbh}{T})\bigg)^{-1} \,.
\end{equation}
The right panel of \fig~\ref{fig:BHProdRate} shows how the fractional error depends on the ratio of plasma temperature to black hole mass. As long as $\Mbh \gtrsim 20 T$, the error on the black hole production rate is less than $10\%$. For black hole production the black hole mass will always be larger than $\Mmin$ which is typically close to $\Mpl$ and the plasma temperature is always below the Planck scale. This makes the low-T limit a viable approximation.

The second approximation we use is that the black hole evaporation is independent of any particle's mass, which is true as long as all species are produced relativistically, i.e. $m_i\ll \Tbh$. This results in the evaporation rate being proportional to $\Tbh^2$ and the particle number production rate being proportional to $\Tbh$. Without this assumption, the constants $\alpha_i$ and $\beta_i$ would be replaced by functions that depend on $m_i/\Tbh$. The validity of our results only start to break down when the dark matter mass comes within an order one factor of the Planck scale. 

Finally, in the main text where present the number density of dark matter at the time of production, $\ndm$, to be
\begin{equation}\label{supp_eq:ndmTRH}
n_{dm}(\Treh)\approx \frac{3\sqrt{\frac{5}{2}} g_*(\Treh)^{3/2} g_{dm} \beta_{dm}}{\pi^3\alpha_{\rm tot}} \frac{\Mmin^8}{\Mpl^5} f_{dm}\bigg(\frac{T_{RH}}{\Mmin}\bigg),
\end{equation}
where the dependence on $\Treh$ is encoded by the function $f_{dm}(x)$. 
We present that to lowest order in $\Treh/\Mmin$, $f_{dm}(x)$ can be approximated as
\begin{equation}\label{supp_eq:ndmTRHFirstOorder}
f_{dm}\bigg(\frac{T_{RH}}{\Mmin}\bigg) \simeq    \left(\frac{\Treh}{\Mmin}\right)^{3/2} e^{-\Mmin/\Treh} \,. 
\end{equation}
However, performing the integrals analytically produces the less intuitive full expression:
\begin{equation} \label{supp_eq:fdmFull}
    f_{dm}\bigg(\frac{T_{RH}}{\Mmin}\bigg) =  \frac{1}{8}\bigg[\bigg(\frac{\Treh}{\Mmin}\bigg)^8 \Gamma\left(\frac{17}{2}, \frac{\Mmin}{\Treh} \right) - \sqrt{\pi}\left({ 1 - \rm erf}\left(\sqrt{\frac{\Mmin}{\Treh}}\right)  \right)\bigg] \,.
\end{equation}
which we have used to produce the results in the main text. Here, $\Gamma(s,x)$ is the upper incomplete Gamma function and $\mathrm{erf}(x)$ is the error function. 

\begin{figure}[htb]
    \centering
    \includegraphics[width=0.48\textwidth]{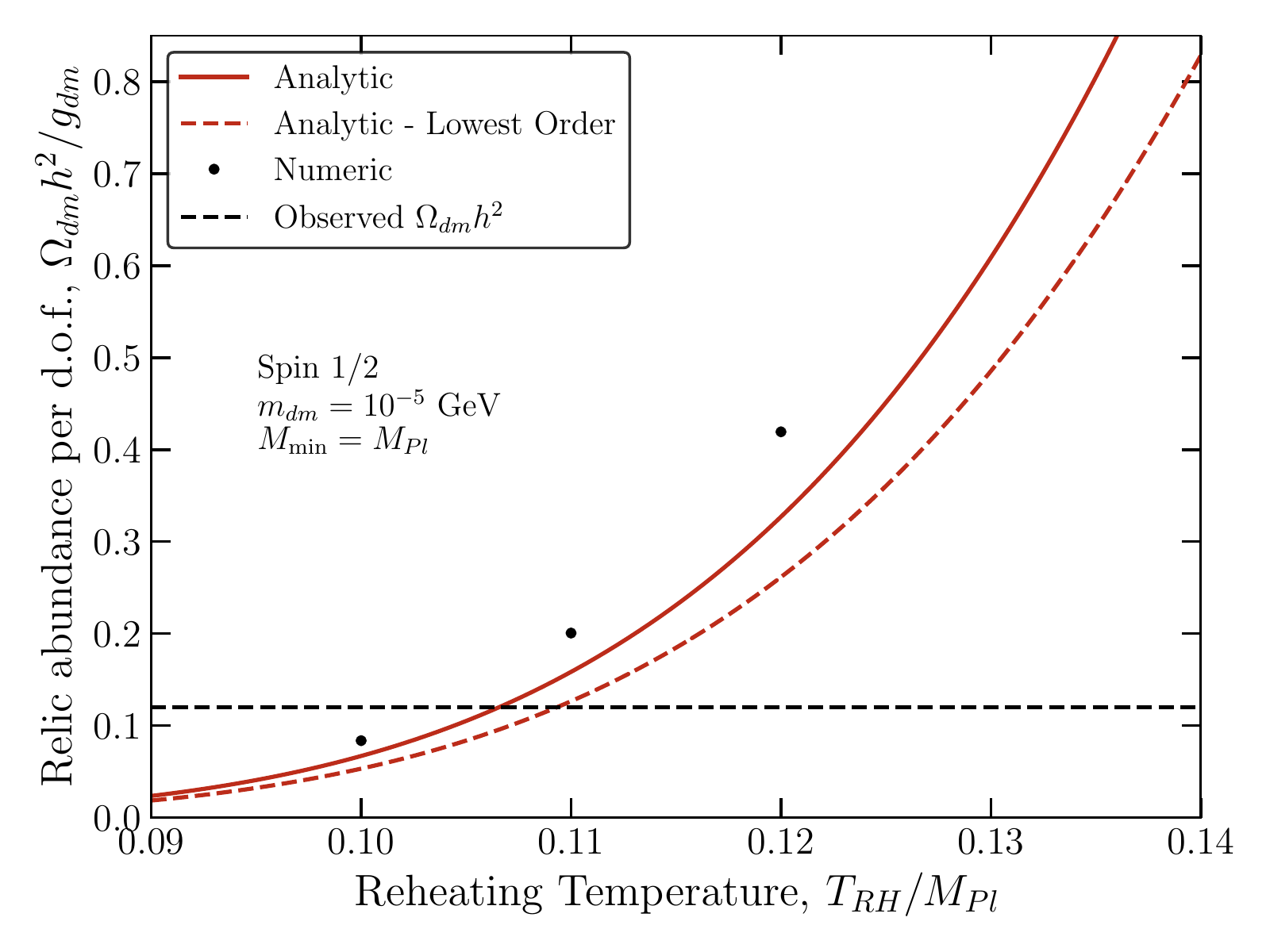}
    \includegraphics[width=0.48\textwidth]{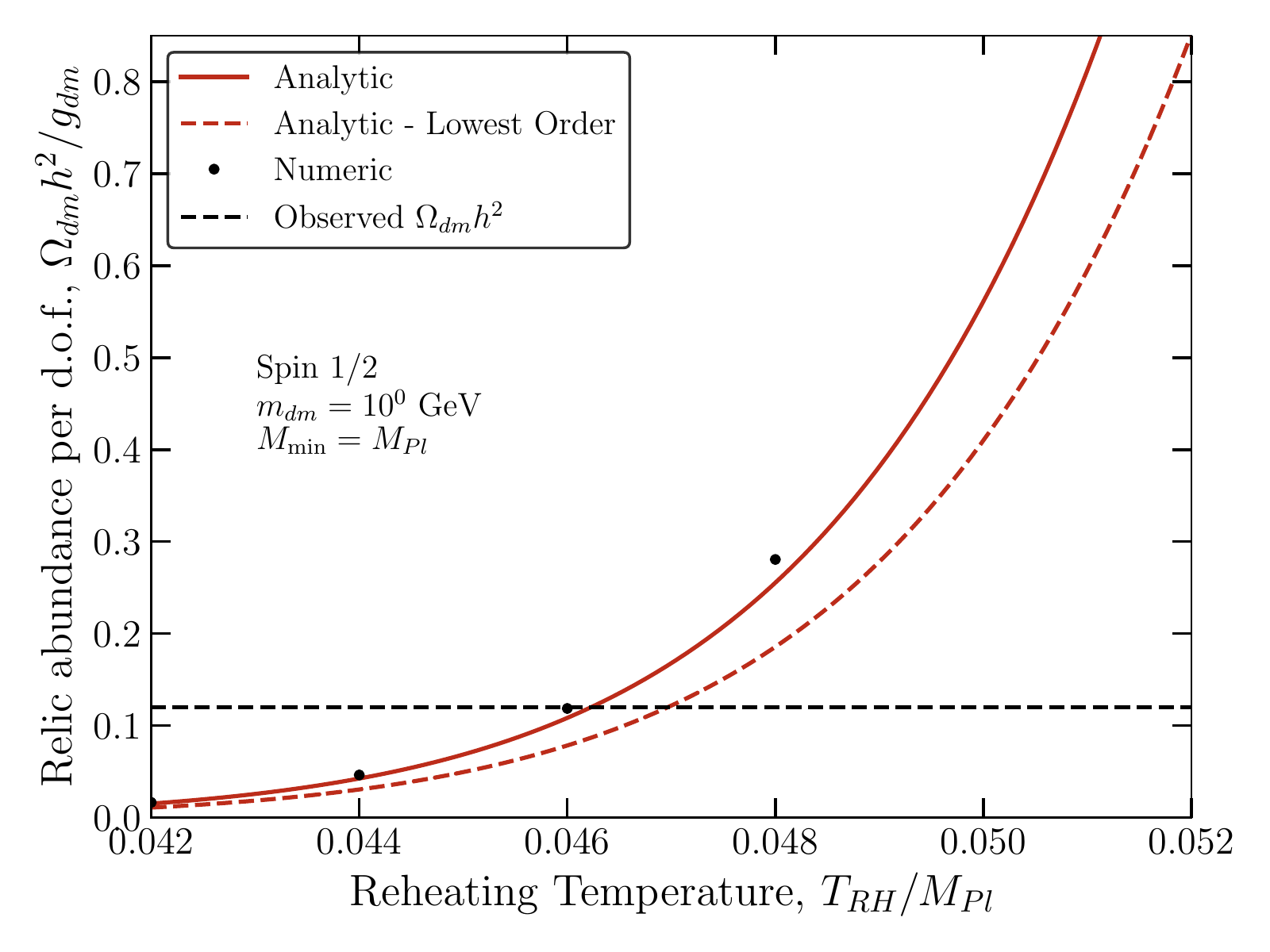}
    \includegraphics[width=0.48\textwidth]{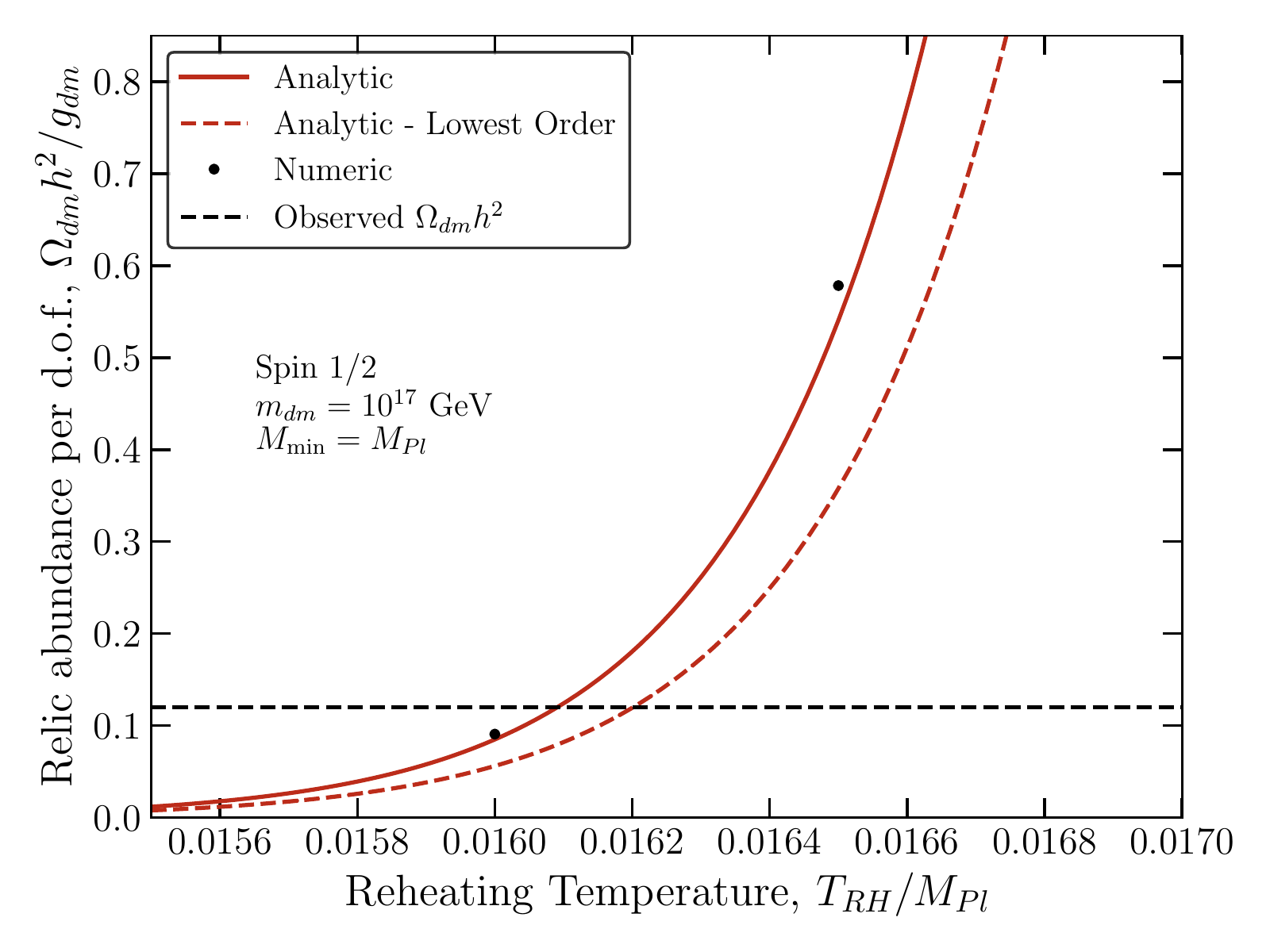}
    \includegraphics[width=0.48\textwidth]{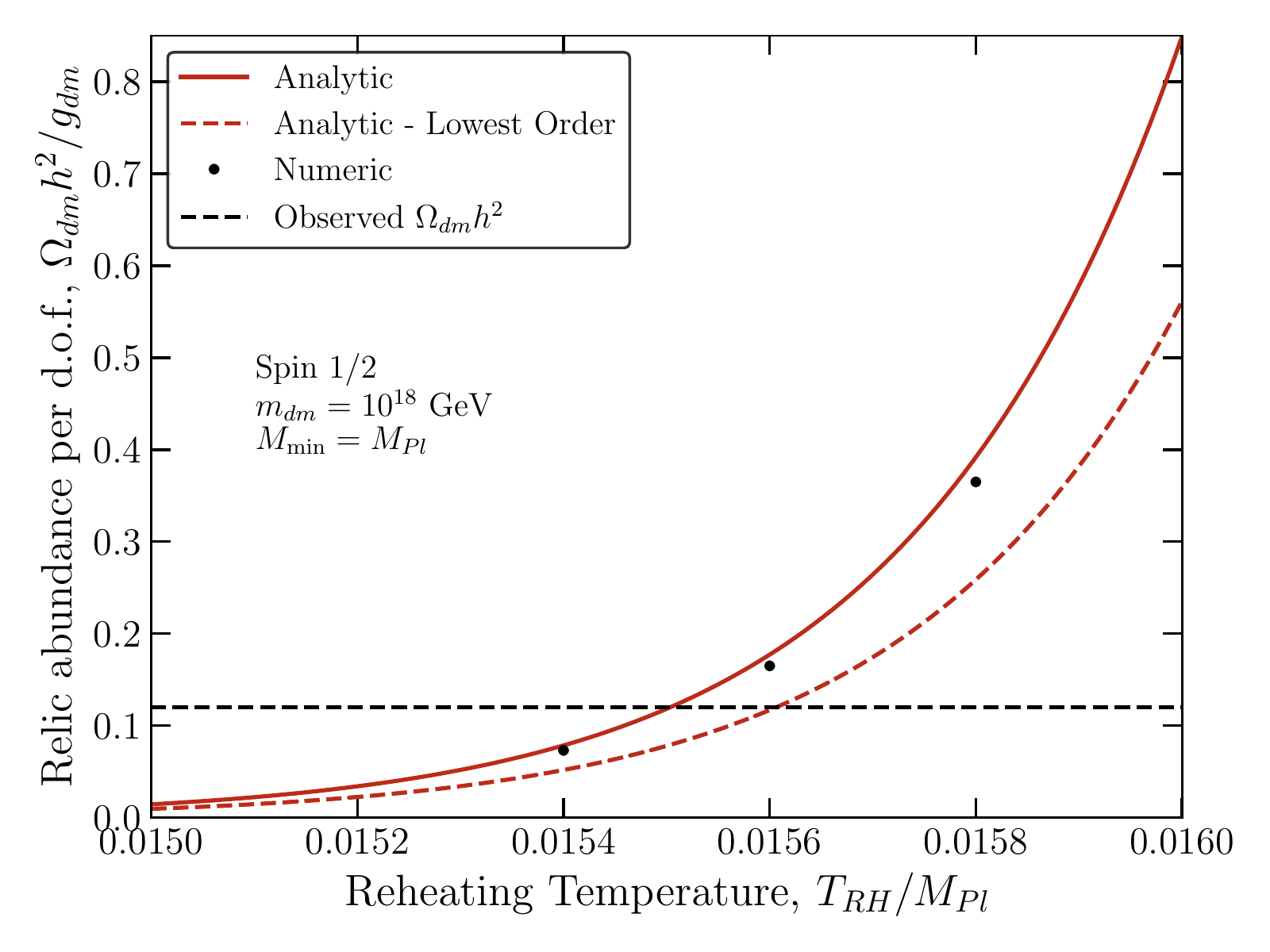}
    \caption{Relic abundance per degree of freedom for fermionic dark. Solid red curve is the relic abundance using the full analytic approximation with $\Treh$ dependence described in \eq~\eqref{supp_eq:fdmFull}. The dashed red curves use the lowest order analytic expression expressed in \eq~\eqref{supp_eq:ndmTRHFirstOorder}. The black points show the true relic abundance found by numerically solving equations without the small $T/\Mbh$ or small $m_{dm}/\Tbh$ approximations. Each panel shows the relic abundance for a different value of $m_{dm}$, zooming in on the $\Treh$ values which produce the observed dark matter abundance. For all calculations, the plasma begins as only Standard Model contents so that $g_*(\Treh)=g_{*s}(\Treh)=106.75$.}
    \label{fig:OmDMZoomIn}
\end{figure}

\fig~\ref{fig:OmDMZoomIn} compares three methods for determining relic abundance of dark matter from evaporating black holes: the analytic expression using \eq~\eqref{supp_eq:fdmFull} used to produce the main text's results is depicted as solid red curves, the lowest order analytic expression using \eq~\eqref{supp_eq:ndmTRHFirstOorder} is depicted as dashed red curves, and the most accurate result found by numerically solving all integrals without the small $\Treh/\Mmin$ or small $m_{dm}/\Tbh$ approximations. In all cases the lowest order analytic approximation underestimates the relic abundance while maintaining the same qualitative dependence on dark matter mass and reheating temperature. These three methods are compared for largest $\Treh$ and $m_{dm}$ values shown in the main text.

The upper left panel of \fig~\ref{fig:OmDMZoomIn} shows the relic abundance of fermionic dark matter with $m_{dm} = 10^{-5}$~GeV. The reheating temperature required to produce the observed relic abundance is $\sim 0.1 \Mpl$. As shown in \fig~\ref{fig:BHProdRate}, the small $T/\Mbh$ approximation under-predicts the number of black holes that form in the early Universe. This in turn leads to under-predicting the relic abundance of dark matter produced by those black holes evaporating. For this set of parameters where $\Treh \sim 0.1 \Mpl$, the relic abundance is underestimated by about $20\%$. However, the upper right panel shows that this error is significantly reduced to below $10\%$ for lower reheating temperatures, $\Treh \sim 0.05 \Mpl$, as is needed to produce the correct relic abundance for fermionic dark matter with $m_{dm} = 1$~GeV.

The lower two panels of \fig~\ref{fig:OmDMZoomIn} demonstrate the success of the analytic approximation even for very heavy dark matter. As discussed above, when $m_{dm} \gtrsim \Tbh$, the analytic approximation ignores the Boltzmann suppression and predicts more dark matter production than the more correct numeric calculation. The lower left panel shows that for $m_{dm} = 10^{17}$~GeV, the analytic approximation does not over-predict the dark matter abundance at all. For even heavier dark matter with a mass of $m_{dm} = 10^{18}$~GeV, the analytic approximation does start to over-predict the relic abundance of dark matter producing an error of about $7\%$. This is the upper limit of dark matter masses produced from black hole evaporation presented in the main text.

\section{Planckeon abundance} \label{app:Planckeon}
If stable Planckeons are indeed the end result of black hole evaporation then one Planckeon would be produced for each black hole. The number density of Planckeons can be found using the same approach as for black hole evaporation, but  using $N_{dm} = 1$ so that
\begin{equation} \label{supp_eq:ndmImplicit}
    n_\mathrm{pl}(\Treh) \approx \int_{\Treh}^0 \!\!\!dT\bigg(\frac{dT}{dt}\bigg)^{-1}  \!\!\int_{\Mmin}^\infty d\Mbh \dfrac{d\Gamma}{d\Mbh} \,.
\end{equation}
Following the same procedure as for dark matter from evaporation, this gives a number density of Plankeons:
\begin{align} \label{supp_eq:nplanckTRH}
   n_\mathrm{pl}(\Treh)&\approx \frac{3\sqrt{\frac{5}{2}} g_*(\Treh)^{3/2} }{4\pi^4} \frac{\Mmin^6}{\Mpl^3}f_{pl}\left(\frac{\Treh}{\Mmin}\right),
\end{align}
where the $\Treh$ dependence is encoded in the function $f_{pl}(x)$. For the full analytic dependence on $\Treh$ we obtained:
\begin{equation}
    f_{pl}\bigg(\frac{T_{RH}}{\Mmin}\bigg) =  \frac{1}{6}\bigg[\bigg(\frac{\Treh}{\Mmin}\bigg)^6 \Gamma\left(\frac{13}{2}, \frac{\Mmin}{\Treh} \right) - \sqrt{\pi}\left({ 1 - \rm erf}\left(\sqrt{\frac{\Mmin}{\Treh}}\right)  \right)\bigg] \,.
\end{equation}
While this differs from the black hole evaporation case provided in \eq~\eqref{supp_eq:fdmFull}, to lowest order the $\Treh$ dependence is the same for Planckeons and dark matter from evaporation. Using $f_{pl} \approx f_{dm}$, we obtain:
\begin{equation}
    n_\mathrm{pl}(\Treh) \approx \frac{3\sqrt{\frac{5}{2}} g_*(\Treh)^{3/2} }{4\pi^4} \frac{\Mmin^6}{\Mpl^3}\left(\frac{\Treh}{\Mmin}\right)^{3/2} e^{-\Mmin/\Treh} \,.
\end{equation}
The results of this expression are shown in purple in the main text.

\end{document}